\documentclass[p3,sort&compress]{elsarticle}
\usepackage[T1]{fontenc}
\usepackage[latin9]{inputenc}
\usepackage{geometry}
\usepackage{color}
\usepackage{amsmath}
\usepackage{amssymb}
\usepackage{graphicx}
\usepackage{esint}
\usepackage[unicode=true,
 bookmarks=true,bookmarksnumbered=false,bookmarksopen=false,
 breaklinks=false,pdfborder={0 0 1},backref=false,colorlinks=true]
 {hyperref}
\usepackage{breakurl}
\begin{document}

\begin{frontmatter}{}

\title{Steady state current fluctuations and dynamical control in a nonequilibrium single-site Bose-Hubbard system}


\author{Xu-Min Chen$^{1}$, Chen Wang$^{1,*}$and Ke-Wei Sun$^{1}$}

\address{
$^{1}$ Department of Physics, Hangzhou Dianzi University, Hangzhou,
Zhejiang 310018, China}

\begin{abstract}
We investigate nonequilibrium energy transfer in a single-site Bose-Hubbard model coupled to two thermal baths.
By including a quantum kinetic equation combined with full counting statistics, we investigate the steady state energy flux and noise power. The influence of the nonlinear Bose-Hubbard interaction on the transfer behaviors is analyzed,
and the nonmonotonic features are clearly exhibited.
Particularly, in the strong on-site repulsion limit, the results become identical with the nonequilibrium spin-boson model.
We also extend the quantum kinetic equation to study the geometric-phase-induced energy pump.
An interesting reversal behavior is unraveled by enhancing the Bose-Hubbard repulsion strength.
\end{abstract}



\end{frontmatter}



\section{Introduction}
Far-from-equilibrium transport, out of linear-response and quasi-equilibrium regimes, has been attracting much attention,
ranging from molecular electronics, quantum magnets to
strongly-correlated materials~\cite{tprosen2008prl,maratner2013nt,shild2014prl,tprosen2014prl,jeisert2015np},
which is of great importance both for fundamental research and practical application.
According to the second law of thermodynamics, energy will flow naturally from the hot source to the cold drain,
under the thermodynamic bias.
Thus, how to control energy transfer in low-dimensional quantum systems becomes a crucial issue, to
unravel the nonequilibrium mechanism and improve the design of efficient devices~\cite{ydubi2011rmp}.

Many proposals have been carried out to study nonequilibrium transport in fermionic systems~\cite{apjauho2007}.
Various interesting phenomena have been unraveled mainly due to the interplay between the voltage and the temperature bias.
In particular, the photovoltaic effect, driven by the nonequilibrium light-electron interaction, provides an efficient way to convert the sunlight to electricity for useful performance~\cite{EB1839cr}.
And the influence of quantum coherence on improving the photovoltaic efficiency was recently proposed~\cite{keforman2013pans,crxu2013prb,dxu2016njp}.
While the thermoelectric effect, one typical kind of heat transfer, describes direct conversion of the thermal bias to electric voltage~\cite{apjauho2007,gjeffrey2008nm}. The relationship between the thermoelectric figure of merit and the conversion efficiency
has been quantitatively characterized~\cite{vcd2005prl,hskim2015pnas}.
Moreover, the Kondo effect, an anomalous feature of the conductance in low temperature regime,
describes the scattering of the conduction electrons mediated by a magnetic impurity~\cite{kondo1964ptp,park2002nature}.
However, as an intimate analogy, the corresponding bosonic systems are lack of exploitation.

Recently, due to fast development of photonics and phononics in quantum transport,
the bosonic systems gain significant popularity~\cite{jscao1,jscao2,lfzhang2010prl,nbli2012rmp,lambert2013np,dwwang2013prl,ksaito2013prl,chien2015np,lv2015prl,lv2016prl}.
As a prototype, the nonequilibrium single-site Bose-Hubbard (SSBH) model is introduced to describe the bosonic system-reservoir interaction~\cite{haake1986pra,purkayastha2016pra}.
The nonlinear Bose-Hubbard coupling is found to be crucial to exhibit novel steady state behaviors.
Such an interaction can be realized by  Kerr interaction in quantum optics~\cite{weinberger2008pml},
by tuning the qubit to the dispersive regime in circuit quantum electrodynamics~\cite{yyin2012pra,tliu2016arxiv},
and by Fermi-Pasta-Ulam interaction in phononic lattice~\cite{nbli2012rmp} and dimer~\cite{jthingna2012prb}.
Though the steady state current in SSBH has been analyzed~\cite{purkayastha2016pra},
the corresponding higher cumulants of energy current are much less exploited,
which is important to characterize the transport features~\cite{campisi2011rmp}.

In the present work, we apply a quantum kinetic equation (QKE) to study the full counting statistics of energy current at steady state in SSBH, which is weakly coupled to two bosonic baths.
The influence of the Bose-Hubbard interaction on the energy current and the noise power is systematically analyzed.
Moreover, we extend the QKE to study the geometric-phase-induced energy pump, and make a comparison with the steady state flux.
This paper is organized as follows:
in Sec. II, we introduce the SSBH model and the quantum kinetic equation combined with the counting field.
The steady state population distribution is analytically expressed.
In Sec. III, we study the steady state energy transfer. Energy current, the corresponding rectification and the noise power
are investigated.
In Sec. IV, we focus on the geometric-phase-induced energy pump.
In the final section, we provide a concise conclusion.

\begin{figure}[tbp]
\begin{center}
\includegraphics[scale=0.5]{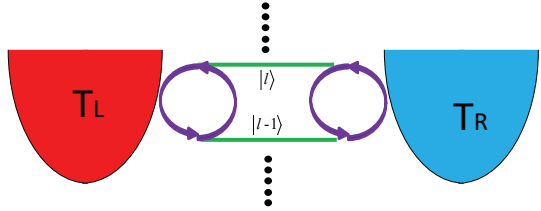}
\vspace{-2.0cm}
\end{center}
\caption{(Color online) The schematic description of a nonequilibrium sing-site Bose-Hubbard system.
The red and blue half pearls are two thermal baths, characterized by  temperatures $T_L$ and $T_R$, respectively;
the central green lines shows the bosonic junction, with $|l{\rangle}$ the occupation state; two purple arrowed circles
demonstrate interactions between the junction and thermal baths.
}~\label{fig0}
\end{figure}

\section{Model and method}
\subsection{Hamiltonian and quantum kinetic equation}
The Model consisting of a bosonic junction coupled to two thermal baths at Fig.~\ref{fig0}, is expressed as
$\hat{H}=\hat{H}_s+\hat{H}_b+\hat{V}_{sb}$.
Specifically, the junction demonstrated as a nonlinear oscillator~\cite{haake1986pra,purkayastha2016pra}, is described as
\begin{eqnarray}~\label{hs}
\hat{H}_s=\omega_0\hat{a}^{\dag}\hat{a}+U\hat{a}^{\dag}\hat{a}\hat{a}^{\dag}\hat{a},
\end{eqnarray}
where $\hat{a}^{\dag}~(\hat{a})$ creates (annihilates) one boson with frequency $\omega_0$,
and $U$ is the onsite boson-boson repulsion strength.
The eigen-solution can be obtained as $\hat{H}_s|n{\rangle}=E_n|n{\rangle}$,
with the eigenvalue
$E_n=(\omega_0n+Un^2)~(n{\ge}0)$,
and  $|n{\rangle}$ the corresponding eigenvector.
Two thermal baths are described as
$\hat{H}_b=\sum_{v=L,R}\hat{H}_{v}=\sum_{v,k}\omega_k\hat{b}^{\dag}_{k,v}\hat{b}_{k,v}$,
where $\hat{b}^{\dag}_{k,v}~(\hat{b}_{k,v})$ is the creating (annihilating) operator of phonons with frequency $\omega_k$ in the $v$th bath.
The junction-bath interaction is given by
\begin{eqnarray}
\hat{V}_{sb}=\sum_{k,v}(g_{k,v}\hat{b}^{\dag}_{k,v}\hat{a}+g^{*}_{k,v}\hat{a}^{\dag}\hat{b}_{k,v}),
\end{eqnarray}
which obeys the particle conservation, with $g_{k,v}$ is the interacting strength.
In this paper, we select $\omega_0$ as the energy unit for simplicity.

Assuming the system-bath coupling strength is weak, we perturb the interaction $\hat{V}_{sb}$ up to the second order.
Based on the Born-Marov approximation, the density operator of the whole system can be approximately decoupled as
$\hat{\rho}(t)=\hat{\rho}_s(t){\otimes}\hat{\rho}_b$, with $\hat{\rho}_s(t)$ the density operator of the junction.
thermal baths are fully thermalized as $\hat{\rho}_b=e^{-\sum_v\beta_v\hat{H}_{v}}/\textrm{Tr}_b\{e^{-\sum_v\beta_v\hat{H}_{v}}\}$,
with the inverse of temperature $\beta_v=1/k_BT_v$.
Hence, the quantum kinetic equation (QKE) of the junction is obtained as~\cite{uweissbook}
\begin{eqnarray}~\label{pe1}
\frac{d}{dt}P_l(t)
&=&\sum_v[-(\kappa^{+}_{v,l+1}+\kappa^{-}_{v,l})P_{l}(t)\nonumber\\
&&+\kappa^{-}_{v,l+1}P_{l+1}(t)+\kappa^{+}_{v,l}P_{l-1}(t)],
\end{eqnarray}
where the population is $P_l(t)={\langle}l|\hat{\rho}_s(t)|l{\rangle}$.
The transition rates in the $v$th bath are
\begin{eqnarray}~\label{rate1}
\kappa^{+}_{l,v}&=&J_v(\Delta_{l})n_v(\Delta_{l})l,\\
\kappa^{-}_{l,v}&=&J_v(\Delta_{l})(n_v(\Delta_{l})+1)l,\nonumber
\end{eqnarray}
where the spectral function is $J_v(\omega)=2\pi\sum_k|g_{k,v}|^2\delta(\omega-\omega_k)$,
the energy gap between $|l{\rangle}$ and $|l-1{\rangle}$ is
\begin{eqnarray}~\label{egap}
\Delta_l=E_l-E_{l-1}=\omega_0+(2l-1)U~(l{\ge}1),
\end{eqnarray}
and the Bose-Einstein distribution function is $n_v(\omega)=1/[\exp(\beta_v\omega)-1]$.
From the dynamical picture, the rate $k^{+}_{l,v}$ describes that one boson is excited from the state $|l-1{\rangle}$ to $|l{\rangle}$ by absorbing one phonon from the $v$th bath;
whereas the rate $k^{-}_{l,v}$ demonstrates that one bosons is released from $|l{\rangle}$  to $|l-1{\rangle}$ by emitting one phonon to the $v$th bath.

In the present work, we specify the spectral function as the typical Ohmic case $J_v(\omega)=\Gamma_v{\omega}e^{-\omega/\omega_{c,v}}$,
with the coupling coefficient $\Gamma_v$ and the cutoff frequency $\omega_{c,v}$.
The Ohmic bath has been widely applied to describe the molecular heat transport~\cite{dsegal2014pre,etaylor2015prl,dsegal2016arpc}, Kondo physics~\cite{ksaito2013prl}, quantum dissipative dynamics~\cite{ajleggett1987rmp,tvorrath2005prl,lwduan2013jcp,uweissbook}
and light-harvesting processes~\cite{lapachon2012pccp,mmohsenibook}.
Here, the cutoff frequency is considered very large, i.e. $\omega_{c,v}{\gg}\omega_0,U,k_BT_v$.
Hence, the bottom part of the bath spectrum mainly contributes to the energy transfer.
The transition rates can be simplified as
$k^{+}_{l,v}=\Gamma_v{\Delta_l}n_v(\Delta_l)l$
and
$k^{-}_{l,v}=\Gamma_v{\Delta_l}(n_v(\Delta_l)+1)l$,
where the factor $e^{-\Delta_l/\omega_{c,v}}{\approx}1$ .

\begin{figure}[tbp]
\begin{center}
\includegraphics[scale=0.55]{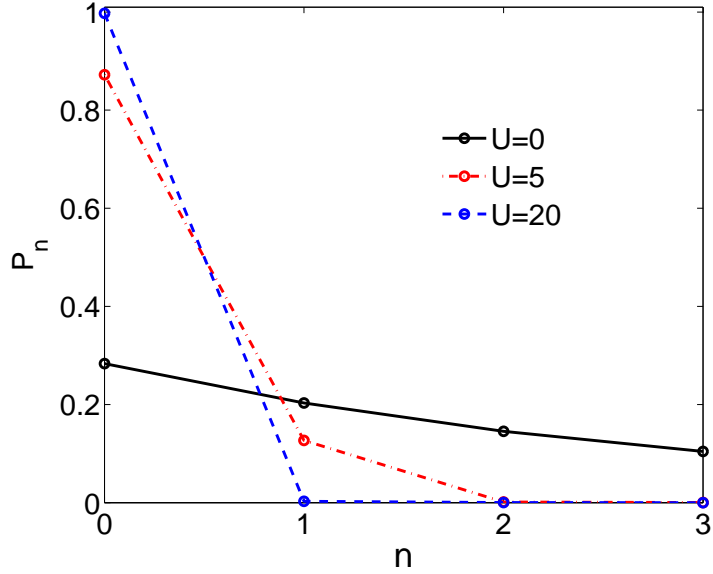}
\end{center}
\caption{(Color online) Steady state population distribution $P_n$ with various onsite boson-boson repulsion strength $U$.
The other parameters are given by $T_L=4$, $T_R=2$ and $\Gamma_L=\Gamma_R=0.1$.
}~\label{fig1}
\end{figure}

\subsection{Steady state}
After a long time evolution, the bosonic junction is completely thermalized ($\frac{d}{dt}P_l(t)=0$),
which results in the balanced relation
$k^{-}_lP_l=k^{+}_lP_{l-1}$.
Thus, the population at level $|l{\rangle}$ can be analytically obtained as~\cite{mvogl2011aop,mvogl2012pra}
\begin{eqnarray}~\label{pop1}
P_l=P_0\prod^l_{m=1}\frac{\sum_v\Gamma_vn_v(\Delta_{m})}{\sum_v\Gamma_v[n_v(\Delta_{m})+1]},
\end{eqnarray}
with the population of the ground state
\begin{eqnarray}
P_0=(1+\sum^{\infty}_{m=1}\prod^{m}_{m^{\prime}=1}
\frac{\sum_{v}\Gamma_vn_v(\Delta_{m^{\prime}})}
{\sum_{v}\Gamma_v[n_v(\Delta_{m^{\prime}})+1]})^{-1}.
\end{eqnarray}
We show the steady state population distribution at Fig.~\ref{fig1}.
In absence of the nonlinearity ($U=0$), the steady state population of the bosonic junction exhibits monotonic decrease.
As the onsite repulsion is turned on, the populations at high energy levels are dramatically suppressed.
Particularly in the strong repulsion regime (e.g., $U=5$), only populations of two lowest energy level states are finite, expressed as
\begin{eqnarray}~\label{state1}
P_0&=&\frac{\sum_{v}\Gamma_v[n_v(\omega_0+U)+1]}{\sum_{v}\Gamma_v[2n_v(\omega_0+U)+1]},\\
P_1&=&\frac{\sum_{v}\Gamma_vn_v(\omega_0+U)}{\sum_{v}\Gamma_v[2n_v(\omega_0+U)+1]},\nonumber
\end{eqnarray}
which becomes identical with the seminal nonequilibrium spin-boson model~\cite{dsegal2005prl}.


\subsection{QKE combined with counting field}
To analyze the steady state energy flux in the single-site Bose-Hubbard system under temperature bias,
we include the full counting statistics to count the energy flow in the right bath~\cite{mesposito2009rmp}.
The Hamiltonian is modified as
$\hat{H}_{\chi}=e^{i\hat{H}_R\chi/2}\hat{H}e^{-i\hat{H}_R\chi/2}=\hat{H}_s+\hat{H}_b+V^{\chi}_{sb}$,
where  the modified junction-bath interaction becomes
\begin{eqnarray}
V^{\chi}_{sb}=\sum_{k,v}(g_{k,v}e^{i\omega_k\chi/2\delta_{v,R}}\hat{b}^{\dag}_{k,v}\hat{a}+g^{*}_{k,v}e^{-i\omega_k\chi/2\delta_{v,R}}\hat{a}^{\dag}\hat{b}_{k,v}),
\end{eqnarray}
with $\delta_{R,R}=1$ and $\delta_{L,R}=0$.
Similarly, based on the Born-Markov approximation at Eq.~(\ref{pe1}), we apply the second order perturbation theory to obtain the modified quantum kinetic equation as
\begin{eqnarray}~\label{qke2}
\frac{d}{dt}P^{\chi}_l&=&-(\kappa^{+}_{l+1}+\kappa^{-}_{l})P^{\chi}_l
+\kappa^{-}_{l+1}(\chi)P^{\chi}_{l+1}\nonumber\\
&&+\kappa^{+}_{l}(\chi)P^{\chi}_{l-1},
\end{eqnarray}
where $P^{\chi}_l={\langle}l|\hat{\rho}^{\chi}_s(t)|l{\rangle}$,
and the modified rates are
\begin{eqnarray}
\kappa^{+}_{l}(\chi)&=&\sum_vJ_v(\Delta_{l})n_v(\Delta_{l})le^{-i\Delta_l\chi_v\delta_{v,R}},\\
\kappa^{-}_{l}(\chi)&=&\sum_vJ_v(\Delta_{l})(n_v(\Delta_{l})+1)le^{i\Delta_l\chi_v\delta_{v,R}}.\nonumber
\end{eqnarray}
In absence of the counting field ($\chi=0$), they return back to the standard transition rate
($\kappa^{\pm}_{l,v}=\kappa^{\pm}_{l,v}(\chi=0)$) at Eq.~(\ref{rate1}).

Next, for discussion convenience, we re-express the kinetic equation at Eq.~(\ref{qke2}) as
\begin{eqnarray}~\label{liouvillian}
\frac{d}{dt}|P_{\chi}{\rangle}=\mathcal{L}_{\chi}|P_{\chi}{\rangle},
\end{eqnarray}
with the vector
$|P_{\chi}{\rangle}=[P^{\chi}_0,P^{\chi}_1,\cdots]^{T}$, and the evoluting matrix $\mathcal{L}_{\chi}$ composed by the modified transition rates.
Then, the generating function can be obtained as
$G_{\chi}(t)={\langle}\textrm{I}|e^{L_{\chi}t}|P_{\chi}(0){\rangle}$~\cite{mesposito2009rmp},
with the unit vector ${\langle}\textrm{I}|=[1,1,\cdots]$, and $|P_{\chi}(0){\rangle}$ the initial population state.
In the long time limit, the generating function can be approximately given by
$G_{\chi}(t){\approx}e^{-\lambda_0(\chi)t}$, where $\lambda_0(\chi)$ is the eigenvalue of $\mathcal{L}_{\chi}$
owing the maximal real part, and $\lambda_0(\chi=0)=0$.
Consequently, the cumulant generating function at steady state is obtained as
$\mathcal{Z}(\chi)=\lim_{t{\rightarrow}\infty}{G}_{\chi}(t)/t=\lambda_0(\chi)$.
And the $n$th cumulant of the energy flux can be expressed as
\begin{eqnarray}~\label{cumulant}
J^{(n)}=\frac{{\partial}^n\lambda_0(\chi)}{{\partial}(i\chi)^n}|_{\chi=0}.
\end{eqnarray}
Particularly, the energy flux is the lowest cumulant case
\begin{eqnarray}~\label{flux}
J=\frac{{\partial}\lambda_0(\chi)}{{\partial}(i\chi)}|_{\chi=0},
\end{eqnarray}
and the noise power is the second lowest cumulant case
\begin{eqnarray}~\label{power}
J^{(2)}=\frac{{\partial}^2\lambda_0(\chi)}{{\partial}(i\chi)^2}|_{\chi=0}.
\end{eqnarray}

\begin{figure}[tbp]
\begin{center}
\includegraphics[scale=0.32]{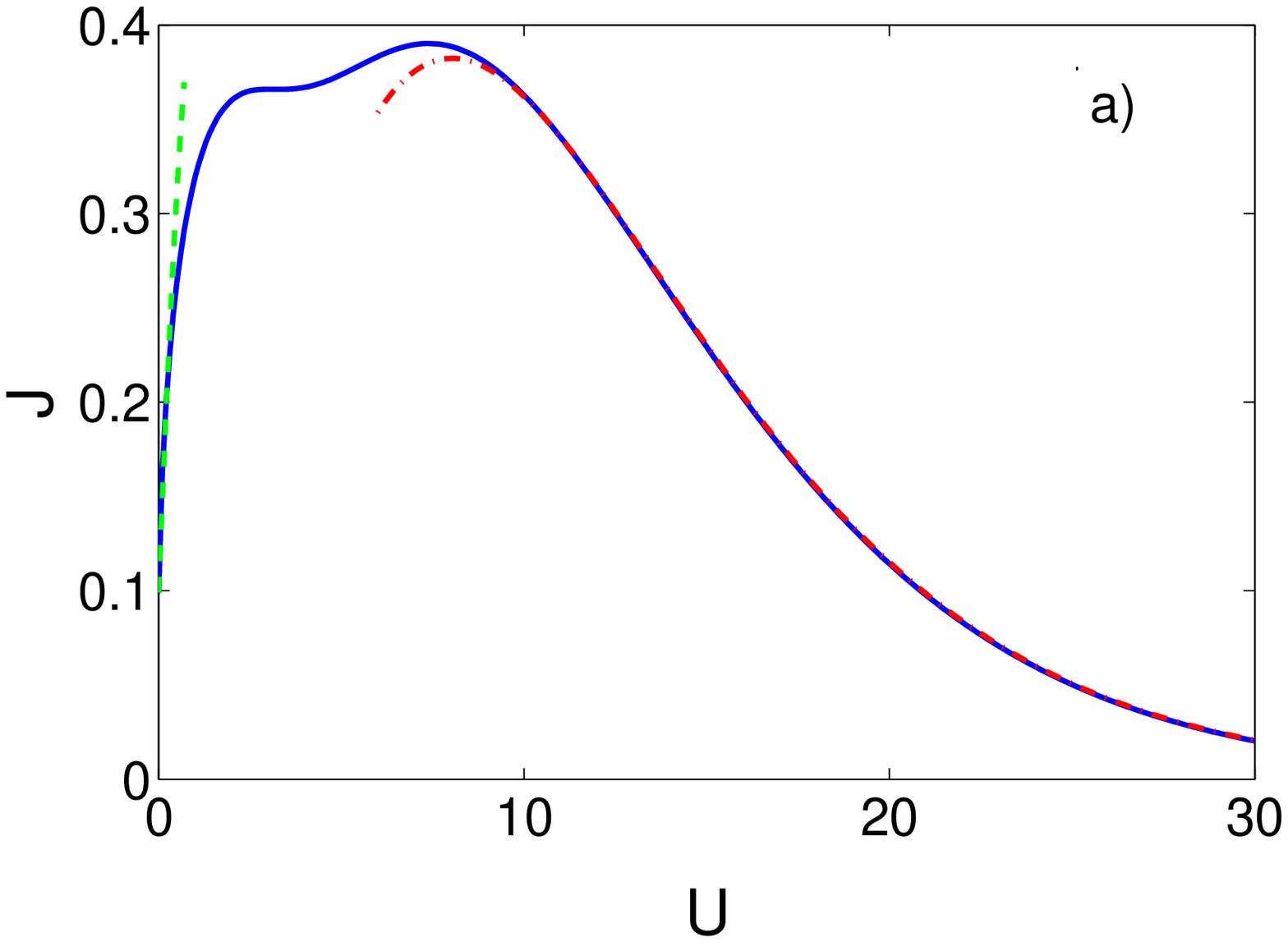}
\end{center}
\begin{center}
\includegraphics[scale=0.4]{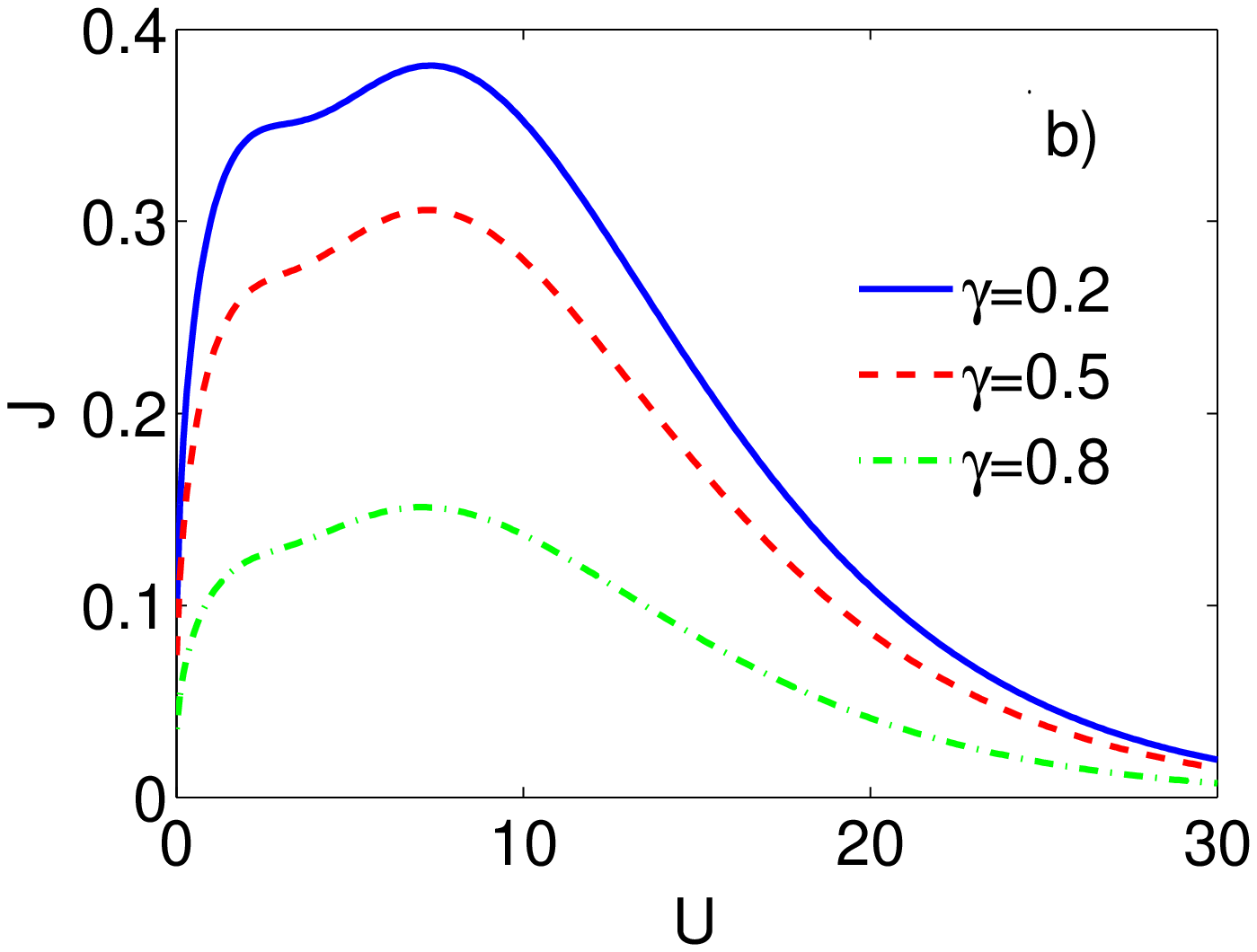}
\end{center}
\caption{(Color online) a) Energy flux in Ohmic baths by tuning onsite repulsion strength $U$ with $\Gamma_L=\Gamma_R=0.1$,
the green dashed line from the result at Eq.~(\ref{jmf}) and the red dashed-dotted line from the result at Eq.~(\ref{j2});
b) energy flux with asymmetric system-bath coupling strength with $\Gamma_L=0.1{\times}(1-\gamma)$ and $\Gamma_R=0.1{\times}(1+\gamma)$.
The other parameters are given by $T_L=4$ and $T_R=2$.
}~\label{fig2}
\end{figure}

\begin{figure}[tbp]
\begin{center}
\includegraphics[scale=0.45]{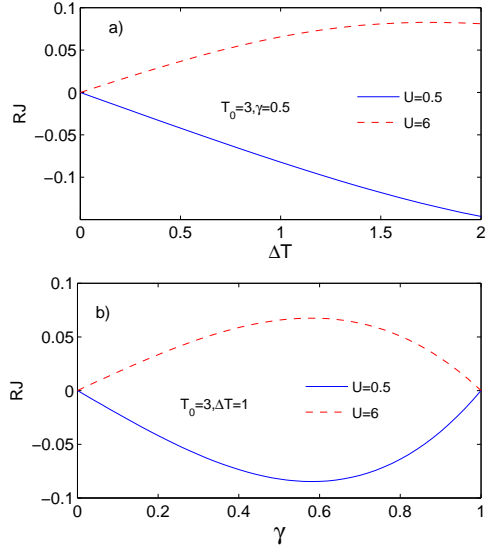}
\end{center}
\caption{(Color online) Energy rectification of the steady state flux ($\textrm{RJ}$) by tuning:
(a) temperature bias between thermal baths;
(b) asymmetric junction-bath coupling factor.
The other parameters are given by $T_L=T_0+{\Delta}T$, $T_R=T_0-{\Delta}T$, $\Gamma_L=0.1{\times}(1-\gamma)$
and $\Gamma_R=0.1{\times}(1+\gamma)$.
}~\label{fig3}
\end{figure}

\section{Steady state energy transfer}
In this section, we study the steady state energy flux, the energy rectification and the noise power.

\subsection{Energy flux}
Based on the eigen solution $\mathcal{L}_{\chi}|P_{\chi}{\rangle}=\lambda_0(\chi)|P_{\chi}{\rangle}$,
the energy flux can be alternatively obtained as
\begin{eqnarray}~\label{j1}
J&=&{\langle}I|\frac{{\partial}\mathcal{L}_{\chi}}{{\partial}(i\chi)}|_{\chi=0}|P_{\chi=0}{\rangle}\\
&=&\sum^{\infty}_{l=1}\Delta_l(\kappa^{-}_{l,R}P_l-\kappa^{+}_{l,R}P_{l-1}),\nonumber
\end{eqnarray}
with $\kappa^{\pm}_{l,R}$ and $P_{l}$ given at Eq.~(\ref{rate1}) and Eq.~(\ref{pop1}), respectively.


We firstly study the influence of the onsite repulsion on the steady state energy transfer with Ohmic baths at Fig.~\ref{fig2}(a).
In absence of repulsion strength ($U=0$), the energy gap at Eq.~(\ref{egap}) becomes level independent as $\Delta_l=\omega_0$,
and the transition rates at Eq.~(\ref{rate1}) are simplified to
$\kappa^{+}_{l,v}=J_v(\omega_0)n_v(\omega_0)l$ and $\kappa^{-}_{l,v}=J_v(\omega_0)(n_v(\omega_0)+1)l$.
Hence, the expression of heat current for the ballsitic transfer is given by
$J=\omega_0^2\frac{\Gamma_L\Gamma_R}{\Gamma_L+\Gamma_R}[n_L(\omega_0)-n_R(\omega_0)]$.
Then, we tune on the repulsion strength.
In the weak repulsion regime, it is found that energy flux is enhanced by increasing the onsite repulsion strength, shown at Eq.~\ref{fig2}(a).
To give an analytical picture, we include the mean-field approximation to simplify boson-boson repulsion at Eq.~(\ref{hs}) as
$\hat{a}^{\dag}\hat{a}\hat{a}^{\dag}\hat{a}{\approx}2n_s\hat{a}^{\dag}\hat{a}-n^2_s$,
with $n_s={\langle}\hat{a}^{\dag}\hat{a}{\rangle}{\approx}[\Gamma_Ln_L(\omega_0)+\Gamma_Rn_R(\omega_0)]/(\Gamma_L+\Gamma_R)$ obtained in the limit of $U=0$.
It results in the mean-field version of the system Hamiltonian
\begin{eqnarray}
\overline{\hat{H}}_s=(\omega_0+2Un_s)\hat{a}^{\dag}\hat{a}-Un^2_s.
\end{eqnarray}
After this treatment, the  Hamiltonian becomes quadratic. Hence, the expression of heat current can be directly obtained as
\begin{eqnarray}~\label{jmf}
J=({\omega}_0+2Un_s)^2\frac{\Gamma_L\Gamma_R}{\Gamma_L+\Gamma_R}
[n_L(\omega_0+2Un_s)-n_R(\omega_0+2Un_s)],
\end{eqnarray}
which clearly exhibits the monotonic behavior by increasing the repulsion strength $U$ (dahsed green line at Fig.~\ref{fig2}(a)).

While at the intermediate repulsion regime (e.g., $U=5$), it is known that high energy state populations have already been dramatically suppressed, shown at Fig.~\ref{fig1}.
Hence, we may include two lowest states to approximately analyze the behavior of the flux in the moderate/strong repulsion regimes.
The expression of the energy flux is given by
$J=(\omega_0+U)^2\Gamma_R[(n_R(\omega_0+U)+1)P_1-n_R(\omega_0+U)P_0]$.
Combined with Eq.~(\ref{state1}), it can be specified as
\begin{eqnarray}~\label{j2}
J=(\omega_0+U)^2\Gamma_L\Gamma_R\frac{n_L(\omega_0+U)-n_R(\omega_0+U)}{\sum_v\Gamma_v[2n_v(\omega_0+U)+1]}.
\end{eqnarray}
The energy flux (red dashed line)  is found to exhibits clearly non-monotonic behavior, shown at Fig.~\ref{fig2}(a).
Hence, we conclude the monotonic behavior of the heat current can be observed under the influence of the onsite boson-boson repulsion.

Then, we study the asymmetric effect of the junction-bath coupling strength on the behavior of energy flux at Fig.~\ref{fig2}(b)~\cite{purkayastha2016pra}.
It is found that by increasing the asymmetric factor $\gamma$, the energy flux is monotonically suppressed.
We conclude that the choice of identical junction-bath coupling strengthes is helpful to strengthen the steady state energy flux.

\subsection{Rectification}
We analyze the energy flux rectification, defined as
\begin{eqnarray}~\label{rj1}
RJ=\frac{|J(\Delta{T},\gamma)|-|J(-\Delta{T},\gamma)|}{|J(\Delta{T},\gamma=0)|},
\end{eqnarray}
with $J(\Delta{T},\gamma)=\frac{\partial{\lambda_0(\chi)}}{{\partial}(i\chi)}|_{\chi=0}$, and ${\Delta}T=T_L-T_R$,
which describes an effect that energy transfer is preferred in one dimension over the opposite~\cite{bli2004prl}.

We firstly study the influence of the temperature bias (${\Delta}T$) on the rectification  at Fig.~{\ref{fig3}}(a).
In the weak repulsion regime (e.g., $U=0.5$), the rectification is negatively amplified by enlarging the temperature bias.
On the contrary, for strong repulsion strength (e.g., $U=6$), the rectification is changed to be positively enhanced.
Then, we analyze the asymmetric effect of system-bath interactions on the rectification at Fig.~{\ref{fig3}}(b).
It is found that for the weak repulsion strength, the rectification exhibits a global valley with the modulated asymmetric factor.
While in the  strong repulsion regime, the rectification exhibits a positive peak.
Particularly, the rectification can be expressed as
\begin{eqnarray}
RJ&=&\frac{4\Gamma_L\Gamma_R[n_L+n_R+1]}{\Gamma_0[\sum_v\Gamma_v(2n_v+1)][\sum_v\Gamma_v(2n_{\bar{v}}+1)]}
{\times}\nonumber\\
&&(\Gamma_L-\Gamma_R)(n_R-n_L),
\end{eqnarray}
It is clearly exhibited that  $(\Gamma_L-\Gamma_R)(n_R-n_L){\neq}0$ results in the emergence of the positive rectification~\cite{dsegal2009prl}.
Moreover, as the asymmetric factor approaches $1$, the rectification effect disappears regardless of the  on-site repulsion strength.
It is known that in this limiting case, one system-bath coupling strength becomes zero, e.g., $\Gamma_L=0$ in the Fig.~\ref{fig3}.
Hence, there is no steady state currents as defined in Eq.~(\ref{rj1}) ($J(\pm\Delta{T},\gamma=1)=0$),
which results in $RJ=0$.

\begin{figure}[tbp]
\begin{center}
\includegraphics[scale=0.45]{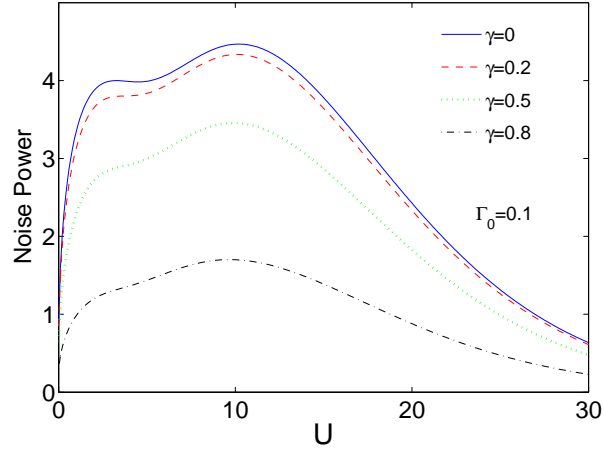}
\end{center}
\caption{(Color online) Relation of noise power with the repulsion strength,
with various asymmetric junction-bath coupling factor.
The other parameters are given by $T_L=4$, $T_R=2$, $\Gamma_L=\Gamma_0{\times}(1-\gamma)$ and
$\Gamma_R=\Gamma_0{\times}(1+\gamma)$.
}~\label{fig4}
\end{figure}

\subsection{Noise power}
We study the influence of the onsite repulsion strength on the noise power,
which is defined as $J^{(2)}={\partial}^2\lambda_0(\chi)/{\partial}(i\chi)^2|_{\chi=0}$ from Eq.~(\ref{power}).
Under symmetric junction-bath coupling condition ($\Gamma_L=\Gamma_R$),
it is found that noise power becomes robust in the weak and mediate coupling regimes,
whereas it is strongly suppressed in the strong repulsion regime.
The main reason is quite similar to the energy flux, that the large energy gap significantly blocks
the state transition between nearest-neighboring states.
Then, we tune on the asymmetric coupling factor $\gamma$.
The noise power is monotonically suppressed by the factor, clearly shown in Fig.~\ref{fig4}.
Therefore, the asymmetric coupling condition deteriorates the generation of the noise power.

\begin{figure}[tbp]
\begin{center}
\includegraphics[scale=0.5]{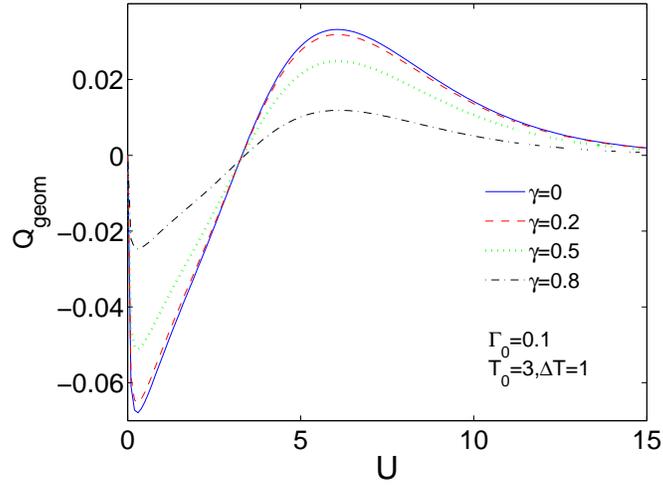}
\end{center}
\caption{(Color online) Influence of the repulsion strength on  geometric-phase-induced energy pump $Q_{geom}=T_p{\times}J_g$,
with various asymmetric junction-bath coupling factor.
The other parameters are given by $T_L=T_0+{\Delta}T\cos\Omega\tau$, $T_R=T_0+{\Delta}T\sin\Omega\tau$, $\Gamma_L=\Gamma_0{\times}(1-\gamma)$ and
$\Gamma_R=\Gamma_0{\times}(1+\gamma)$.
}~\label{fig5}
\end{figure}
\section{Adiabatic energy pump}
The time-dependent control of heat transfer in junction systems has attracted dramatic attention~\cite{nbli2012rmp,jren2010prl,jren2012prl}.
In particular, the nonlinear effect is revealed to be crucial to exploit the adiabatic quantum pump.
Therefore, we apply the geoemtric-phase-like modulation to study the influence of repulsion strength on the energy transfer
in the single-site Bose-Hubbard model.

As the bosonic junction is adiabatically modulated by external fields
(e.g., $T_{L(R)}(t)$ in two baths), the evoluting matrix becomes time dependent $\mathcal{L}_{\chi}(t)$.
Then, the generating function is composed by two components as~\cite{jren2010prl,nasinitsyn2007epl}
\begin{eqnarray}
\lim_{t{\rightarrow}\infty}=e^{G_{\chi}(t)}=\exp([G_{d}(\chi)+G_{g}(\chi)]t),
\end{eqnarray}
where $G_{\chi}$ is the cumulant generating function.
$G_{d}(\chi)=-\frac{1}{T_p}\int^{T_p}_0dt\lambda_0(\chi,t)$ is the dynamical contribution,
with $T_p$ the modulating period.
It straightforwardly results in the dynamical energy flux
$J_{dyn}=\frac{\partial}{{\partial}(i\chi)}G_d(\chi)|_{\chi=0}$.
In absence of the external modulation, it reduces to the steady state flux from Eq.~(\ref{flux}).
$G_{g}(\chi)$ is the geometric contribution, which is expressed as
\begin{eqnarray}
G_{g}(\chi)=-\frac{1}{T_p}\int^{T_p}_0dt
{\langle}\phi_{\chi}(t)|\frac{\partial}{{\partial}t}|\psi_{\chi}(t){\rangle},
\end{eqnarray}
where $|\psi_{\chi}(t)~({\langle}\phi_{\chi}(t)|)$ is the right (left) eigenvector, corresponding to $\lambda_0(\chi,t)$.
Assuming the driving fields are expressed as $u_1(t)$ and $u_2(t)$,
the geometric phase induced energy flux is specified as~\cite{mvberry1984prsla,dxiao2010rmp}
\begin{eqnarray}
G_{g}(\chi)=-\frac{1}{T_p}{\oint}(du_1{\langle}\phi_{\chi}|\frac{\partial}{{\partial}u_1}|\psi_{\chi}{\rangle}
+du_2{\langle}\phi_{\chi}|\frac{\partial}{{\partial}u_2}|\psi_{\chi}{\rangle}).
\end{eqnarray}
According to the Stocks theorem, it can be re-expressed as~\cite{jren2010prl,jren2012prl,cwang2017pra}
\begin{eqnarray}
G_{g}(\chi)=-\frac{1}{T_p}\int\int_{u_1,u_2}du_1du_2\mathcal{F}_{\chi}(u_1,u_2),
\end{eqnarray}
with the Berry-like curvature
\begin{eqnarray}
\mathcal{F}_{\chi}(u_1,u_2)={\langle}\partial_{u_1}\phi_{\chi}|\partial_{u_2}\psi_{\chi}{\rangle}-
{\langle}\partial_{u_2}\phi_{\chi}|\partial_{u_1}\psi_{\chi}{\rangle}.
\end{eqnarray}
Then, the geometric-phase-induced current is given by
\begin{eqnarray}
J_{g}&=&\frac{{\partial}G_g(\chi)}{{\partial}(i\chi)}|_{\chi=0}\\
&=&-\frac{1}{T_p}\int\int_{u_1,u_2}du_1du_2\frac{\partial}{{\partial}(i\chi)}\mathcal{F}_{\chi}(u_1,u_2)|_{\chi=0}.\nonumber
\end{eqnarray}

We firstly study the geometric-phase-induced energy pump under the symmetric junction-bath coupling condition ($\gamma=0$) at Fig.~\ref{fig5}.
In absence of the on-site repulsion ($U=0$), there is no geometric-phase-induced pump, due to the fully harmonic behavior of the molecular junction.
From the previous works, it is proposed that nonlinearity of the system is crucial to exhibit the geometric contribution of the flux~\cite{jren2010prl,jren2012prl}.
Hence, the present result is consistent with the proposal.
Moreover, in the weak repulsion regime the energy pump shows negative enhancement.
It is in sharp contrast to the behavior of the steady state flux at Fig.~\ref{fig2}(a), which shows positive robustness.
As is known that steady state flux reflects the behavior of quasi-energy spectrum of the Liouvillian matrix at Eq.~(\ref{liouvillian}),
whereas geometric flux is sensitive to  modulations of the corresponding eigen-states.
Therefore, due to different information they capture, the influence of the repulsion strength on the geometric-phase-induced energy flux is significantly distinct from the counterpart
for the steady state flux.
Then, it rises quickly and reaches the globally positive peak with intermediate repulsion strength.
As the repulsion strength further increases, the energy pump is suppressed and asymptotically approaches zero.
Next, we tune on the junction-bath coupling asymmetric factor. It is found that the energy pump is monotonically suppressed by enlarging the asymmetric factor.
Hence, we conclude that the asymmetric factor plays a similar role in both the steady state and geometric energy flux.

\section{Conclusion}
In summary, we study the quantum energy transfer in a single-site bosonic junction weakly coupled to two thermal baths with temperature bias.
The quantum kinetic equation combined with the full counting statistics, is included to analyze the energy current and corresponding current fluctuations (e.g., noise power).
Steady state population distribution is analytically obtained.
Particularly, in strong onsite boson-boson repulsion regime, two lowest energy level states mainly contribute to steady state behaviors.
Without the adiabatic modulation, in weak onsite boson-boson repulsion regime, steady state energy flux with Ohmic baths is enhanced by increasing the replusion stregnth.
It is analytically explained  based on the mean-field scheme.
Then, the current shows maximal with intermediate repulsion strength,
whereas it asymptotically approaches to zero in the strong repulsion limit, which is similar to the behavior of the nonequilibrium spin-boson model.
Moreover, we study the energy flux rectification by tuning the temperature bias and asymmetric coupling factor.
It is found that the behavior of energy rectification with weak repulsion strength is different from the counterpart in the strong repulsion regime.
With the adiabatic modulation, the geometric-phase-induced energy flux shows significant distinction from the steady state counterpart.
Specifically, the geometric-phase-induced energy flux shows negative enhancement in the weak onsite repulsion regime.
Moreover, it shows reversal behavior and becomes positive maximal with intermediate repulsion strength.
As the repulsion strength further enlarges, the geometric-phase-induced energy flux is suppressed and gradually decays to zero.

In previous works~\cite{jswang2008epjb,hnli2008epjb}, the nonequilibrium green function approach was exploited to investigate the quantum heat transfer in the anharmonic boson junction
with strong junction-bath interaction, which dramatically affects the behavior of heat current and high order fluctuations.
Hence, we may include this novel method to analyze the influence of interplay between strong junction-bath coupling
and on-site boson-boson interaction
on the heat transfer.

\section{Acknowledgements}
C.W. is supported by the National Natural Science Foundation of China under Grant No.11704093, No. 11547124 and No. 11574052,
and K.W.S. is supported by the National Natural Science Foundation of China under Grant No. 11404084.

$^{*}$ Corresponding author. Email:wangchenyifang@gmail.com

\end{document}